# Comment on "Extended Halogen Bonding between Fully Fluorinated Aromatic Molecules: Kawai *et al.*, ACS Nano, 2015, 9, 2574–2583"

Arpita Varadwaj,[†,‡] Pradeep R. Varadwaj, *,[†,‡] Helder M. Marques,[¶] Koichi Yamashita[†,‡]

[†]Department of Chemical System Engineering, School of Engineering, The University of Tokyo 7-3-1, Hongo, Bunkyo-ku, Japan 113-8656, [‡]CREST-JST, 7 Gobancho, Chiyoda-ku, Tokyo, Japan 102-0076, [¶]Molecular Sciences Institute, School of Chemistry, University of the Witwatersrand, Johannesburg, 2050 South Africa

Kawai and co-workers have advanced a number of ideas on fluorine-centered non-covalent bonding formed between the fully-fluorinated aromatic compounds phenyleneethynylene (1,2-bis[2,3,5,6-tetrafluoro-4-[(2,3,4,5,6-pentafluorophenyl)ethynyl]phenyl]ethyne (BPEPE-F18).[1] The study has rapidly attracted considerable attention because it displayed attraction between negative sites, leading to unusual bonding topologies that were essential for the design of ordered supramolecular architectures on a metal surface. An interaction of this type is recently shown not just limited to the fluorine atom in molecules, but is also extended to other halogen derivative in other molecules that have assisted in the design materials of similar kind.[2] The authors of the study[1] have claimed the intermolecular interactions they identified are observed by them for the first time, but these have in fact been known for decades but were under-appreciated. We offer five comments on the conclusions made by Kawai *et al.*[1], which we believe to be incorrect and misleading.

Before we proceed to comment on their study, we want to clarify the underlying concept of a σ-hole. It is a concept that has explained many phenomena in a wide range of chemical systems. We have received comments from some anonymous reviewers of our work that advance the notion that *a σ-hole is always positive and that it is misleading to say a σ-hole can be negative*. Several reports have appeared which put forward this contention, leading to significant misunderstanding and controversy. Similar statements are found in the paper on which we offer comment,[1] in which the authors have stated that *there is no σ-hole on the fluorine in the chemical systems examined* yet they still form directional interactions.

**What is a σ-hole? Do we understand it?**

Answer: Despite many (thousands) articles published on σ-hole centered halogen bonding since 2007, there has been a great deal of confusion about its underlying concept. This has prompted the originators of the concept to recently revisit the topic in order to reclarify its conceptual foundations.[3] By definition, a *σ-hole* has literally nothing to do with a hole; it refers to a tiny region on the surface of a covalently bonded atom in a molecule, which is deficient of electron density. Through the mapping between isoelectron density and electrostatic potential, some researchers color this region in blue, in green, or even in red, depending on their choice of presentation. This coloring scheme should not and cannot lead to a belief that the σ-hole is always positive (as often assumed). Thus, when we refer to a region as being "electron deficient" along the outer extension of an R–X covalent σ-bond (X = F, Cl, Br, I; R = remainder part of the molecule), we are actually comparing this electron deficiency on the outer portion of atom X with the lateral sides of the same atom in the same molecule. Because there is always a mismatch between charge densities of the axial and lateral sides of atom X, this has been called as an anisotropic distribution of the charge density. Because of this specific feature, researchers have attributed this amphoteric nature of the atom X to what has been termed "a fallacy of net atomic charges".[4-5]

**Is there any way to quantify a σ-hole?**

Answer: A σ-hole on halogen X along the outermost extension of a R–X covalent bond can be positive, negative, or even neutral.[4, 6-10] Conceptual DFT can provide evidence for the existence of a σ-hole on fluorine, as has been discussed elsewhere.[11] The often used formalism is to quantify and visualize this through the molecular electrostatic surface potential (MESP) approach. As such, the nature and strength of σ-hole on atom X has been judged by the *local maximum* of electrostatic surface potential, $V_{s,max}$, along the outermost extension of the R–X bond.[3, 10] Thus, if $V_{s,max} < 0$, one generally comes up with a σ-hole with negative electrostatic potential (which has been referred to as a negative σ-hole); if $V_{s,max} > 0$, one often recognizes this as a σ-hole with positive electrostatic potential (which is what has been found in most of the halogenated hydrocarbon compounds with X = Cl, Br and I and is called a positive σ-hole, or simply a σ-hole, or an electrophilic cap, and so on); and if there is no local maximum of electrostatic potential ($V_{s,max} = 0$), then the σ-hole is absent.[7-8, 12] This accords with the view of Politzer and co-workers, according to whom the term "σ-hole" originally referred to the *electron-deficient outer lobe of a half-filled orbital of predominantly p character* involved in forming a covalent bond. If the *electron deficiency* is *sufficiently large*, a region of positive electrostatic potential results which can interact attractively (non-covalently) with a *negative* site.[10] *The strength of the inter- or intramolecular interaction generally correlates well with the magnitudes of the positive and negative electrostatic potentials of the σ-*hole (we called this positive and negative σ-holes, respectively) and the negative site (such as N in NH$_3$).[9] Positive σ-holes can also be found on covalently-bonded Group IV-VI atoms and can interact electrostatically with a negative site.[3, 9] Nevertheless, it is worth stressing that the concept of a σ-hole has not always been fully understood and has led to contradictory statements such as, for example, the σ-hole in H$_3$C–F, H$_3$C–Cl and CF$_4$ being neutral[13] or negative,[5, 14] and with the notion that the latter two cannot halogen bond. Some of us,[15-16] and others,[17-18] have already proved some of these views to be *perversely counterintuitive* since H$_3$C–Cl and CF$_4$ can halogen bond as both of them feature weakly positive σ-holes, whereas that on H$_3$C–F is entirely negative.

These rationalizations, which are against the views of an anonymous reviewer, are in reasonable agreement with the most recent discussions provided in the paper of Politzer and coworkers where they revisit the concept of a σ-hole.[3] As they write, "a covalently-bound atom typically has a region of *lower electronic density*, a "σ-hole", on the side of the atom opposite to the bond, along its extension", clarifying the confusion that one should not assume a "σ-hole" to be always positive, as some of us have demonstrated previously on several occasions.[7-8, 12] The authors have further clarified that the birth of the term, introduced in a conference in 2005, was initially used to denote the localized positive potentials on the outer sides of many univalent halogen atoms. However, it was soon recognized that it *was more appropriate to use "σ-hole" to describe lower electronic densities* that are found along the extension of σ bonds. They further clarify that "there is usually (*but not always*) a positive electrostatic potential associated with this lower electron density, i.e., a σ-hole, as initially found by Brinck *et al.* for univalent halogens. It is through this positive potential that attractive interactions with negative site occur." They further conclude that "Fluorine, the least polarizable and most electronegative halogen, tends to have least positive potentials; in fact, *the potentials due to its σ-holes are often negative, although less so the surrounding 0.001 a.u. contour. The fluorines then have negative $V_{s,max}$ ... these fluorines do have σ-holes*", a view which is indeed in line with our findings.[7-8, 12]

A positive σ-hole of reasonable strength on X should always be visible on a 0.001 a.u. isoelectron density mapped molecular electrostatic surface.[3, 9-10, 15] If it is very weak and $V_{s,max} < 0$, it does not normally show up because of overlapping with the (very negative) lateral sides.[3, 7-8, 12] In fact, it is a matter of graphical generation and manipulation using suitable constraints and presentation methods, which might assist to visualize a weak σ-hole. Also, we have demonstrated in a few instances that one has to be careful to use an appropriate computational method to glean insight into the nature of the σ-hole, whether it is weakly positive or negative. An inappropriate use of a computational method can otherwise lead to results that can be misleading.[15-16]

On several occasions, the weakly positive or negative $V_{s,max}$ on X along R–X cannot be visualized. *But this certainly does not mean that the σ-hole is absent*, as drawn in the Kawai paper.[1] To clarify this, let us consider Figure 1h and i, in which, the $V_{s,max}$ associated with the σ-holes of the fluorine atoms in $C_6F_5CN$ and $C_6F_4(CN)_2$ are both positive (+0.74 and +4.80 kcal mol$^{-1}$, respectively), but graphically do not show up. Similarly, the σ-hole on the fluorine along the C–F bond extensions in $C_6F_6$ (Figure 1a), in $C_6F_5$–X (Figure 1j (X =NO, l (X = OH) and r (X = SH)) and others are negative, but also do not show up. We attribute this as a shortcoming of the σ-hole theory, causing misunderstanding and misinterpretation of many underlying phenomena such as in the study of Kawai *et al*.[1] The same argument as above applies, for example, to an intermolecular interaction A⋯X, as discussed by Murray and others.[19] In particular, it was shown in that study that once the non-covalent interaction has been formed, the regions of negative and positive potential on A and X that were the driving force for it will not be visible in the final A⋯X complex as they will have been at least partially polarized during the formation of the interaction.[19] This is a qualitative picture. Now the immediate question arises: can this intermolecular interaction be realized quantitatively through this model? The answer is *no*, there is no way to do this with this specific approach because there is no way to calculate the electrostatic potential on the surfaces of the atoms on the monomers in the complex since their surfaces have already overlapped to form the complex, thereby leading to a shortcoming of the MESP model.

**1 – One of the claims in the paper of Kawai *et al.*,[1] which appears in the abstract as well as in a later review by the same author,[20] is that "*fully fluoro-substituted aromatic molecules, on the contrary, are generally believed not to form halogen bonds due to the absence of a σ-hole*".**

Answer: *This view is perversely counterintuitive* since it refers to the traditional understanding of the reactivity of fluorine in molecules. The notion is in sharp contrast with many recent reports (for example[7-8, 12, 15, 21-25]), including some from the originators of the term 'σ-hole',[13, 26] in which F in a variety of fluorinated hydrocarbons does display a σ-hole.

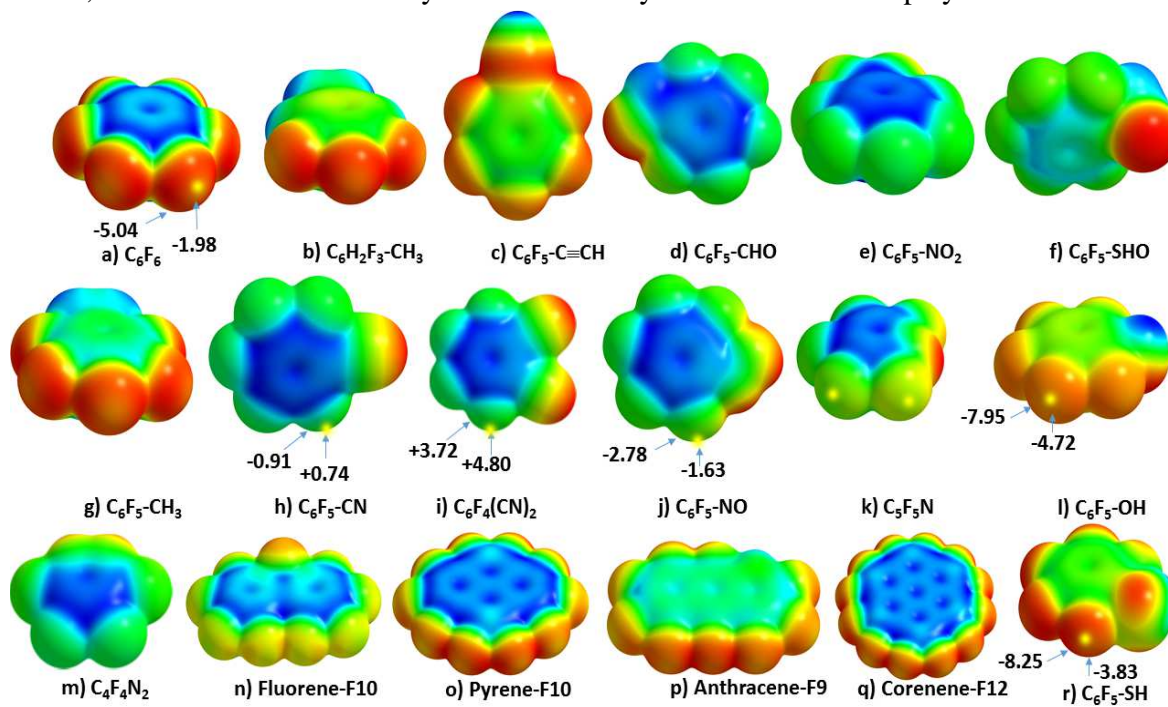

**Figure 1.** PBE/6-311++G(2d,2p) level 0.001 a.u. isodensity mapped molecular electrostatic surface potential of some partially and fully fluorinated aromatic compounds. The maximum and minimum ($V_{s,max}$ and $V_{s,min}$ (in kcal mol$^{-1}$) respectively) of MESP are shown on some fluorine atom in a few of these compounds, demonstrating anisotropy in the charge density. For clarity, yellow dots representing $V_{s,max}$ are shown only for a few cases.

We provide further evidence in support of our argument above by considering a wide range of fully and partially fluoridated aromatic systems, which have been energy-minimized in this study using Gaussian 09[27] based density functional theory. Figure 1 illustrates the MESPs for these compounds. It shows that there are indeed σ-holes on the surfaces of F along the outer extensions of the C–F bonds, which are not visible. These are either positive or negative, which are determined by the negative and positive sign of $V_{s,max}$.

**2 – Kawai and co-workers[1] used high resolution atomic force microscopy to study in-plane F⋯F contacts formed by the fully fluorinated BPEPE-F18 molecule. Figure 2**

depicts the molecule (our result). While they did not report the actual values of electrostatic surface potentials for the axial and lateral sites of the fluorine of the C–F bonds ($V_{s,max}$ and $V_{s,min}$, respectively), they claimed that the structure "contain[s] three C–F⋯F bonds among neighboring molecules. *While the σ-hole is absent, the scheme of the C–F⋯F bonding has a high similarity to halogen bonding.*"

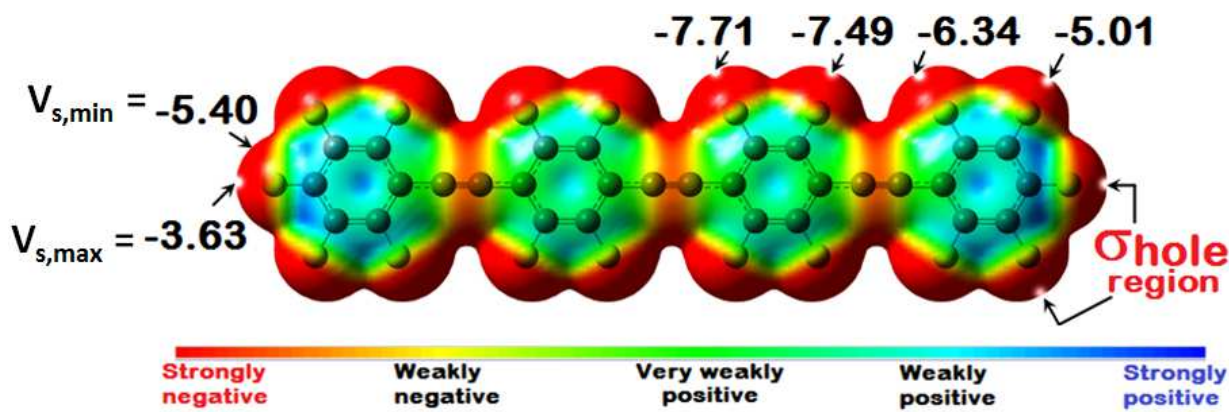

**Figure 2.** M062X/6-311++G(d,p) level 0.001 a.u. isodensity mapped molecular electrostatic surface potential (MESP) of BPEPE-F18. The maximum and minimum ($V_{s,max}$ and $V_{s,min}$, respectively) of MESP around some fluorine atoms in this compound are displayed, showing the presence of a σ-hole with negative MESP ($V_{s,max} < 0$) along the outermost extensions of the C–F bonds.

Answer: As can be seen from Figure 2, the σ-hole on the fluorine in BPEPE-F18 *is not absent*, as claimed by Kawai et al.[1] As already discussed above, it is there, but bears a negative electrostatic potential ($V_{s,max} < 0$). A few of them on the fluorine atoms along C–F bonds are shown; their strengths are not all equal. Because the σ-holes on the fluorine in BPEPE-F18 are negative, and because the lateral sites of the same fluorine atoms are more negative ($V_{s,min} < 0$), the electron density profile of the F atom in BPEPE-F18 is anisotropic. The overlapping between the negative potentials of these sites does not permit the negative σ-hole to become visually distinguished. We further note that the σ-hole on the fluorine in BPEPE-F18 is so negative that its nature cannot be changed by changing the level of correlated theory and basis set. Nevertheless, the aforementioned anisotropic feature is common in halogens in alkylhalides, in which cases, the σ-hole is generally positive (later sides negative) that participates in halogen bonding (other noncovalent interactions).[5, 28-29]

**3 –** The authors have stated that i) "*Here, we present clear evidence that an anisotropically distributed negative electrostatic potential around a fluorine atom in a fluoro-substituted aromatic molecule together with a large dispersion force induces a highly directional bonding and gives rise to a well ordered supramolecular structure on a Ag(111) surface at low temperature. ii) On this basis, we argue that the deep minimum of the energy versus θ curve, i.e., the high directionality of the C–F⋯F bond, originating from the F–F* *electrostatic interactions. iii) The dispersion force takes the major part in the attraction, but the driving force of the directional bonding is due to the anisotropically distributed electrostatic potential around the fluorine atoms."*

Answer: First, the three views i)-iii) do not match with each other. Second, they are also incorrect since the directionality of the F···F interactions do not come from either dispersion (known to be non-directional) or electrostatics. It should be understood as a balance between all kinds of intermolecular interactions arising from electrostatics, exchange, repulsion, polarization and dispersion. In particular, the F···F interactions reported by Kawai et al should be understood as the mutual competition between attraction (polarization and dispersion) and repulsion (electrostatic and exchange repulsion), with the strength of the overall interaction being modulated by dispersion. A notion similar to that of Kawai *et al.* has recently been challenged by Thirman and colleagues,[30] who have argued that the correlation of the binding energy with the negative or positive $V_{s,max}$ associated with the σ-hole does not mean that halogen bonding should be and can be understood solely as electrostatically driven, since insofar as charge transfer is concerned, electrostatic effects play an important role in the stability of the interaction. This interesting study eventually led to the same conclusion as us that the equilibrium geometry of any chemical system should not be understood as a consequence of electrostatics only as argued by Stone,[31] but rather is a balance between electrostatics, Pauli repulsion, charge transfer, polarization, and dispersion, among others, which have all different length scales depending on the nature of reactive sites interacting to form binary and large-scale complexes. These are collectively and simultaneously responsible for determining the overall interaction and the directionality, despite these energy components only emerging from rigorous treatment at different levels of decomposition of the binding energy. These conclusions also address the views of an anonymous reviewer who stated that some of the halogen assisted synthons in the molecular assemblies reported elsewhere[32-33] would arise from the sum of the van der Waals attraction (London dispersion forces) and the electrostatic repulsion between terminating halogen atoms.

Although two types of interaction modes lead to the formation of the "wind-mill" topology the authors report, these were not correctly assigned. We call these interactions formed between the fluorine atoms of interacting monomers to be the consequence of the combination of two types bonding topologies, type-I and type-II halogen-centered non-covalent interactions (Figure 3a). (Herein, we emphasize that halogen bonding and halogen-centered non-covalent interactions are not the same thing). The former type, which is known to be dispersion driven, is typical of non-directional halogen···halogen bonding.[29] The latter is typical of type-II halogen-centered non-covalent fluorine bonding because it is directional, in which case there is a significant HOMO and LUMO overlap as a consequence of donation of electron density to the R–X antibonding orbital $\sigma *$ orbital of the acceptor (an experimental view[29]). We, and others,[29, 34-35] have reported intermolecular interactions of this type elsewhere.[7-8, 12] It should therefore be understood that the type-I and type-II interaction topologies responsible for the development of "windmill" type synthons are a consequence of competition between all sorts of interacting forces. Desiraju et al. classified type-I and type II interactions as symmetrical and bent interactions, respectively.[36]

**4 – The authors of paper have argued that "However, fluorine has the strongest electronegativity and a low polarizability, *so that it usually has no σ-hole*. The σ-hole is strongly affected by the electron attracting power, and so, for instance, the cyano moiety behind the C–F bond in FCN and the oxygen in hypofluorite can induce a σ-hole. However, if the electron extraction is not strong enough, F has an intrinsic negative electrostatic potential. *Simple fluorosubstituted hydrocarbons, such as $C_6F_6$ and $CF_4$, are categorized as the latter case.* Thus, *the halogen bond in such molecules is rarely seen in analyses of crystal structure databases and the stronger C–F⋯π interaction generally dominates*. Therefore, analyses of bulk crystal structures cannot reveal the F⋯F contact. Since, however, the freedom of molecules on a surface is restricted, the in-plane intermolecular interaction at the F⋯F contact can readily be investigated." [This view is reported in the original paper and is repeated elsewhere.[20]]**

Answer: The statements italicized above are incorrect. We argue against this suggestion because no matter how and which way the fluorine is involved to form a molecule, there will always be a local electrostatic potential ($V_{s,max}$) on its surface along the outermost extension of the R–F σ-bond (Figure 1). It (the σ-hole) can be either positive or negative, or even neutral in very rare cases. Introduction of a strongly electron-withdrawing group R such as –CN can only tune the nature of the σ-hole from negative to positive. This enforces the electron deficient σ-hole to form a relatively stronger intermolecular interaction with a negative site.

The assertion that because of its high electronegativity and low polarizability, the F atom in aromatic- and alkyl-fluorides such as $C_6F_6$ and $CF_4$ bears no σ-hole has recently been challenged by us.[15] For instance, we have recently examined both $C_6F_6$[7-8] and $CF_4$[15] and their complexes with several negative sites localized on different molecules. We showed that the σ-holes on the surfaces of the F atom along the C–F bond extensions in $CF_4$ are weakly positive ($V_{s,max}$ = +0.45 kcal mol$^{-1}$ at the MP2/6-311++G(2d,2p) level), a finding which is in sharp contrast with the contention made by Kawai *et al*.[1] Similarly, in two other studies, we have shown that the $V_{s,max}$ associated with the σ-holes on F along the C–F bond extensions in $C_6F_6$ are negative (*viz*. Figure 1a). Indeed, these findings are clearly in sharp contrast with the views of Kawai and coworkers that assumes no σ-holes on F. It is also counter to the opinion of an anonymous reviewer who stated that the F along the C–F bond extensions in $C_6F_6$ will be weakly positive with MP2. We make it clear that no matter what level of theory used, our demonstration[7-8] that the fluorine in $C_6F_6$ will always be entirely negative is always valid, with the potential along the axial sites (σ-holes) always smaller than the lateral sites because the electron densities of the lone-pairs dominate. To prove this, we have energy-minimized $C_6F_6$ with MP2 level of theory in conjunction with two different basis sets 6-31+G(d) and 6-311++G(2d,2p), and then calculated MESPs on both the geometries (as we did this with PBE, Figure 1a). These calculations with MP2 show that the $V_{s,max}$ and $V_{s,min}$ associated with the axial (σ-hole) and lateral sites are respectively –3.50 and –6.51 kcal mol$^{-1}$ with the former basis set, and –3.00 and –5.82 kcal mol$^{-1}$ respectively with the latter basis set.

We comment further that the crystal structure of CF$_4$ is stabilized by various type-I and type-II halogen centered non-covalent interactions as discussed elsewhere.[15] Therefore, the experimental and theoretical results on CF$_4$ are against the claims of Kawai *et al.*, which means that the suggestion that *the halogen bond in such molecules is rarely seen in analyses of crystal structure databases and the stronger C–F···π interaction generally dominates*, is not at all true for CF$_4$.

**5 – Is this kind of C–F···F–C bonding interactions observed by Kawai *et al.* driven by metal surfaces (and packing and solid state effects) as generally believed and as reported in several other papers?**

To answer this most important question, we have carried out gas phase calculations on the dimers, trimers and tetramers of C$_6$F$_6$.[7] Similar calculations were performed for derivatives of pentafluorophenylethynyls (Figure 2). From these gas phase results, it can readily be concluded that the C–F···F–C bonding between the negative sites is not a consequence of solid state and surface effects, as has been contended.[1] It is rather the consequence of the nature of the fluorine in the fully-fluorinated aromatic compounds that govern such unified topologies of bonding interactions, which they mimic the observations in the solid state; they are largely driven by the effects of polarization and dispersion.[7-8] This view is in agreement with Baker *et al.*[37] who have examined the solid state structures of three compounds that contain a perfluorinated chain, CF$_3$(CF$_2$)$_5$CH$_2$CH(CH$_3$)CO$_2$H, CF$_3$(CF$_2$)$_5$(CH$_2$)$_4$(CF$_2$)$_5$CF$_3$ and (CF$_3$(CF$_2$)$_5$CH$_2$CH$_2$)$_3$P═O, and have identified a number of C–F···F–C and C–F···H–C interactions that are closer to the sum of the van der Waals radii of the two fluorine atoms. These authors have undertaken a comprehensive computational chemistry investigation with Quantum Theory of Atoms in Molecules and have confirmed the specific C–F···F–C interactions they identified are certainly not due simply to crystal packing.

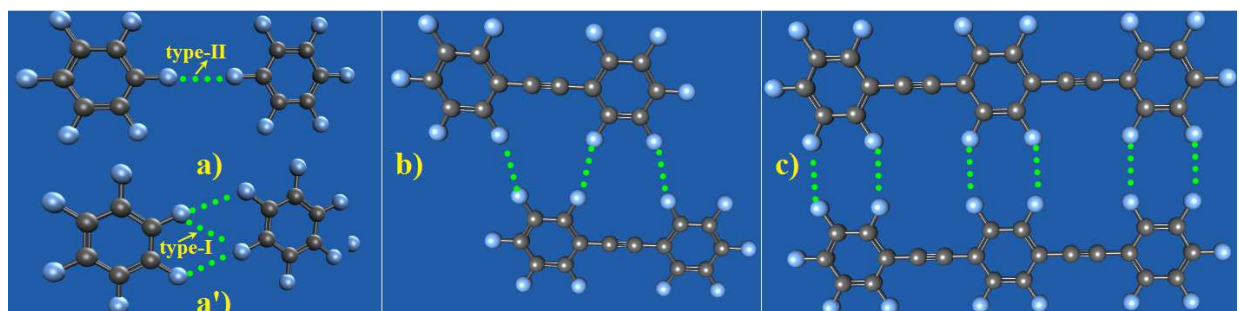

**Figure 3.** Geometries of the (C$_6$F$_6$)$_2$ (a and a′), (C$_{14}$F$_{10}$)$_2$ (b), and (C$_{22}$F$_{14}$)$_2$ (c) binary complexes. The structures in a) and b) were energy-minimized with M06-2X/6-311++G(d,p), while c) is the direct experimental geometry collected from the Cambridge Structural Database (ref codes BEWZEJ and BEWZEJ01).[38] In the first two cases, the F···F type-II directional bonding is clearly evident, while in a′) and c) the occurrence of the type-I topology between the same atoms is not surprising.

As indicated by the dotted lines in Figure 3a, each C–F bond points to a fluorine atom on an adjacent molecule, which results in highly directional C–F···F bonds (~180°). A consequence of this is the appearance of another F···F diagonal interaction, with C–F···F bond angles ~120°, which can be understood as a sort of forced interactions. These directional features are the same as the halogen bonding,[29, 36] since halogen bonding interactions are known to the community as type-I and type-II, in which, the latter are commonly understood as the consequence of a positive site on the halogen and the negative site and the former as dispersion dominant (non-directional). Clearly, despite the geometrical similarity, the various interactions identified by Kawai et al.[1,20] as well as those in Figure 3, cannot and should not be called halogen bonding as this potentially will create significant controversy about the general understanding and definition of halogen bonding.

In all cases, the F···F distance is close to 300 pm, which is slightly larger than twice the van der Waals radius of fluorine (294 pm). These geometrical features indicate that these bonds do not differ from a typical halogen···halogen (type-I) bonds,[29, 36] even though the interacting species forming these interactions are entirely negative.

**AUTHOR INFORMATION**

Corresponding Author

*E-mail: pradeep@tcl.t.u-tokyo.ac.jp

**ACKNOWLEDGMENTS**

AV, PRV and KY thank Institute of Molecular Science, Okazaki, Japan for supercomputing facilities received for all calculations, and thank CREST project for generous funding (Grant No. JPMJCR12C4). HMM thanks the National Research Foundation, Pretoria, South Africa, and the University of the Witwatersrand for funding.